\documentclass[12pt]{iopart}
\usepackage{iopams}
\usepackage{graphics}
\usepackage[next]{inputenc}
\usepackage[dvips]{epsfig}
\def\be{\begin{equation}}
\def\ee{\end{equation}}

\def\bi{\begin{itemize}}
\def\ei{\end{itemize}}
\def\bn{\begin{enumerate}}
\def\en{\end{enumerate}}
\def\bea{\begin{eqnarray}}
\def\eea{\end{eqnarray}}

\def\ba{\begin{array}}
\def\ea{\end{array}}
\def\bd{\begin{displaymath}}
\def\ed{\end{displaymath}}
\def\la{\langle}
\def\ra{\rangle}

\begin{document}

\title[Space modulation on 1D alternating spin-1/2 model]{Effects of a space modulation on the behavior of a 1D alternating Heisenberg spin-1/2 model}

\author{Saeed Mahdavifar$^1$ and Jahanfar Abouie $^{2,3}$\footnote{The authors have the same contributions in the work.}}

\address{$^{1}$Department of Physics, University of
Guilan,41335-1914, Rasht, Iran}
\address{$^{2}$ Department of Physics, Shahrood University of Technology, Shahrood 36199-95161, Iran}
\address{$^{3}$ School of physics, Institute for Research in Fundamental Sciences (IPM),
Tehran 19395-5531, Iran}

\ead{Mahdavifar@guilan.ac.ir}
\ead{jahan@shahroodut.ac.ir}
\date{\today}
\begin{abstract}

\leftskip 2cm \rightskip 2cm

The effects of a magnetic field ($h$) and a space modulation
($\delta$) on the magnetic properties of a one dimensional
antiferromagnetic-ferromagnetic Heisenberg spin-1/2 model have
been studied by means of numerical exact diagonalization of finite
size systems, nonlinear sigma model and bosonization approach. The
space modulation is considered on the antiferromagnetic couplings.
At $\delta=0$, the model is mapped to a gapless L\"{u}ttinger
liquid phase by increasing the magnetic field. However, the space
modulation induces a new gap in the spectrum of the system and the
system experiences different quantum phases which are separated by
four critical fields. By opening the new gap a magnetization
plateau appears at $\frac{1}{2}M_{sat}$. The effects of the
space modulation are reflected in the emergence of a plateau in other physical
functions such as the F-dimer and the bond dimer order parameters, and the pair-wise entanglement.
\end{abstract}


\pacs{75.10.Jm, 75.10.Pq}

\maketitle
\section{Introduction} \label{sec1}
The effects of a magnetic field on the magnetic properties of low-dimensional quantum magnets at zero temperature
has attracted much attentions in recent years.
One of the very interesting phenomena is the appearance of a
magnetization plateau in the magnetization curve.
This behavior can be viewed as an essentially macroscopic quantum phenomena, and
has gained much attentions, recently.
When a plateau is appeared
the energy gap is opened, which can be in some senses regarded as
a kind of generation of the Haldane conjecture \cite{haldane83}.
In a seminal work, M. Oshikawa, M. Yamanaka and I. Affleck
studied the magnetization of a general class of the Heisenberg spin chains at the presence of a magnetic field. They
showed that the plateaus can be appeared when the
magnetization per site $m$ is topologically quantized by
$n(s-m)=integer$, where $s$ is the magnitude of the spin, and $n$
is the period of the ground state determined by the explicit
spatial structure of Hamiltonian \cite{oshikawa97}.

The bond alternating Heisenberg spin-1/2 chains which are obtained by a space
modulation in the exchange couplings \cite{takada92, hida-0-92, hida93,
hida-1-92, kohmoto92, yamanaka93, hida-0-94} are a particular class of the low-dimensional quantum magnets to observe
the magnetization plateau at zero temperature.
The bond alternating Antiferromagnetic-Ferromagnetic spin-1/2 chains have a gap in the
spin excitation spectrum and reveal extremely rich quantum
behaviors in the presence of an external magnetic
field \cite{sakai95, yamamoto05, mahdavifar08, abouie08}. By turning
the magnetic field, the excitation gap is reduced and reach to
zero at the first critical field. Simultaneously, the magnetization start to
increase up to its saturation value, $0.5$ at the second critical
field. More enhancement of the field re-opens the gap and the saturation plateau
appears in the magnetization curve. These models have only two
plateaus at zero and saturation values of the magnetization. It
has been found that a space modulation in
the exchange couplings can affect on the behaviors of the field induced
magnetization. For example the bond alternating
Ferromagnetic-Ferromagnetic-Antiferromagnetic (F-F-AF) trimerized
Heisenberg spin-1/2 chains have exotic behaviors by
changing the magnetic field \cite{ajiro94}. This model can be realized in the
$Cu$-compounds such as $3CuCl_2.2dx$.
The magnetization has a plateau at
$\frac{M}{M_{sat}}=\frac13$, where $M$ and $M_{sat}$ are the
magnetization and its saturated values,
respectively \cite{hida-1-94, okamoto96}. The mid-plateaus have also been
appeared in the bond alternating
Ferromagnetic-Antiferromagnetic-Antiferromagnetic (F-AF-AF)
trimerized Heisenebrg spin-1/2 chains \cite{gu06}.
It has been also shown that the static structure factor dose not vary with the external magnetic field at the plateau state \cite{gong08}.

The other examples are the tetrameric bond-alternating Ferromagnetic-Ferromagnetic-Antiferromagnetic-Antiferromagnetic
(F-F-AF-AF) Heisenberg spin models. A realization for the
spin-1/2 chain with F-F-AF-AF alternations is the compound
$Cu(3-Clpy)_2(N_3)_2(3-Clpy=3-Chloroyridine)$ \cite{escuer98}.
In such a spin-$\frac12$ system, there is a gap from the
singlet ground state to the triplet excited states in the absence
of a magnetic field.
At $\frac{M}{M_{sat}}=\frac12$, a plateau is appeared in the magnetization curve by applying a magnetic field \cite{lu05}.
The temperature dependence of the magnetization, magnetic susceptibility and
specific heat have been also studied in the mid-plateau state of this model by means of
transfer-matrix renormalization group method \cite{gong09}.

\begin{figure}[h]
\centerline{\includegraphics[width=8cm,angle=0]{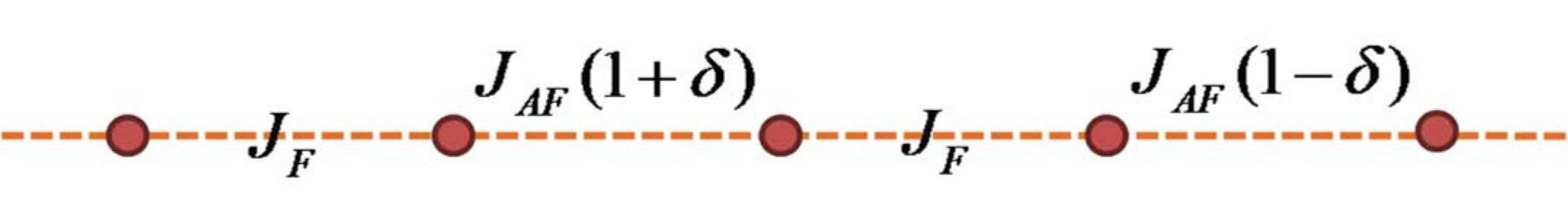}}
\caption{(Color online.) Schematic representation of a tetrameric
spin chain} \label{chain}
\end{figure}

In this paper, by considering a different bond alternating Heisneberg spin-1/2 chain, we study the other physical properties of the mid-plateau state such as the string order and pair-wise entanglement, F-dimer and bond-dimer order parameters.
The Hamiltonian of the model
is
\begin{eqnarray}
\nonumber H&=&-J_F\sum_{j=1}^{N/2}{\bf S}_{2j}\cdot{\bf S}_{2j+1}\\
&+&J_{AF}\sum_{j=1}^{N/2}[1+(-1)^{j} \delta]{\bf S}_{2j-1}\cdot{\bf S}_{2j}-h \sum_{j=1}^{N} S_j^{z},~~~~~~ \label{hamiltoni}
\end{eqnarray}
where ${\bf S}_{j}$ is the spin-$\frac{1}{2}$ operator on the
$j$-th site. $J_F$ and $J_{AF}^{\pm}=J_{AF}(1\pm \delta)$ denote
the ferromagnetic ($J_F>0$) and antiferromagnetic ($J_{AF}>0$) couplings, respectively. $\delta$ is the space modulation parameter
and $h$ is proportional to the external magnetic field. The unit cell of the model has been shown in Fig. (\ref{chain}).
The bond alternating spin-1/2 AF-F chains are good candidates
for studying the L\"{u}ttinger liquid phase.
A suitable realization of the bond alternating AF-F spin chain is
the $\rm{Cu}$ based $\rm{(CH_3)_2NH_2CuCl_3}$ compound.
Linked-cluster calculations and bulk measurements show that
$\rm{DMACuCl_3}$ is also a realization of the spin-1/2 alternating
AF-F chain with nearly the same strength of antiferromagnetic and
ferromagnetic couplings \cite{Ston07}. A space modulation changes the AF couplings alternatively and influences on the quantum properties of the system.

Using the exact diagonalization of a finite size system and employing the conformal field theory it has been already found that for $\delta\neq0$ a magnetization plateau exist at half of the saturation value \cite{Chen98}. Moreover, the width of the mid-plateau is dependent to the values of $\delta$ and the critical exponent for the plateau
width is obtained by the level spectroscopy method.
In the present work the different properties of the model such as F-dimer and bond-dimer order parameters, string order parameter, and the pair wise entanglement of the plateau state are studied.

The outline of the paper is as follows. In the forthcoming section we discuss the model in the
strong AF coupling limit and derive an effective spin chain Hamiltonian. In section III we summarize the results of
the analytical field theory studies. In section
IV, the results of exact diagonalization Lanczos method are presented. Finally, we discuss and summarize
our
results in section V.

\section{Effective Hamiltonian}  \label{sec2}

\begin{figure}
\centerline{\includegraphics[width=8cm,angle=0]{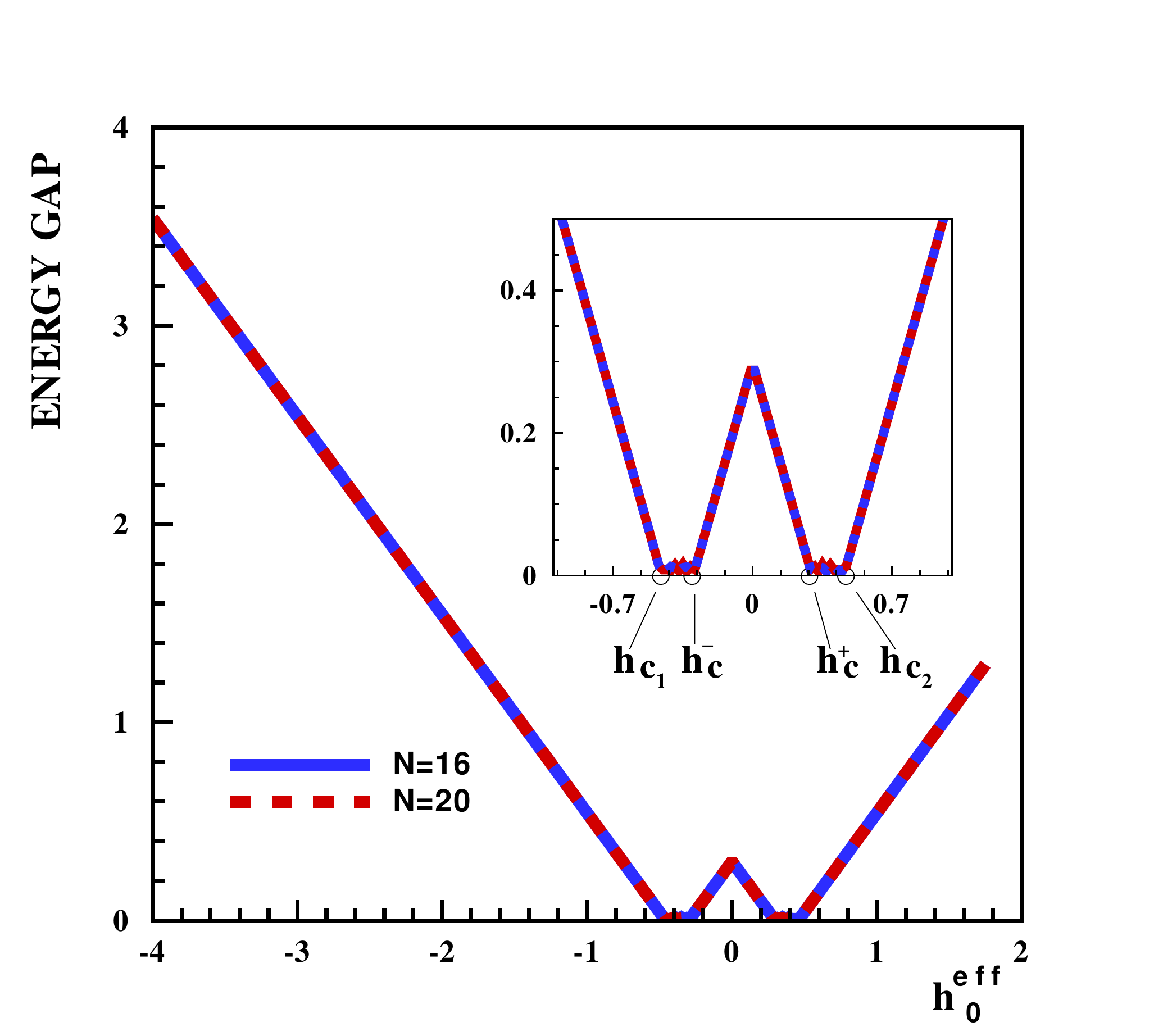}}
\caption{(Color online.) Difference between the two lowest energy levels of the effective Hamiltonian versus the magnetic field $h_0^{eff}$, for chains with different lengths $N=16,20$ and exchange parameters $J_{F}=1.0$, $J_{AF}=9/2$ and $\delta=1/9$.} \label{energy-gap-effective}
\end{figure}

The easiest way to obtain the effective model, which manifestly
displays mechanism for generation of the sequence of new scales is
to start from the limit of noninteracting block of pairs $J_{F}=0$
and strong magnetic field $h\simeq J_{AF}$ \cite{mila98}. In the
limiting case of the strong AF coupling $J_{AF}\gg J_{F}$ and $J_{AF}\gg\delta
J_{AF}$ the model can be mapped to an effective spin chain
Hamiltonian \cite{mahdavifar08}. At $J_{AF}\gg J_{F}$, the
system behaves as a nearly independent block of pairs. Indeed an
individual block of spin pairs may be in a singlet or a triplet
state with the corresponding eigenvalues given by
\begin{eqnarray}
&E^{\pm}(S)&=-\frac{3}{4}J_{AF}^{\pm},\nonumber \\ &E^{\pm}(T_{1})&=\frac{1}{4}J_{AF}^{\pm}-h,~~E^{\pm}(T_{0})=\frac{1}{4}J_{AF}^{\pm},\nonumber \\
&E^{\pm}(T_{-1})&=\frac{1}{4}J_{AF}^{\pm}+h.\label{block-pairs}
\end{eqnarray}
When $h$ is small, the ground state consists of a product of pair
singlets. As the magnetic field $h$ increases the energy of the
triplet state $|T_1\rangle$ decreases and at $h=J_{AF}^{\pm}$
forms together with the singlet state, a doublet of almost
degenerate low energy states, split from the remaining high energy
two triplet states. Thus, for a strong enough magnetic field we
have a situation when the singlet $|S\rangle$  and triplet
$|T_1\rangle$ states create a new effective spin $\tau=1/2$
systems. On the new singlet-triplet subspace and up to a constant,
we easily obtain the effective Hamiltonian
\begin{eqnarray}
H^{eff}&=&-\frac{J_F}{2}\sum_{j=1}^{N/2}[\tau_{j}^{x}\tau_{j+1}^{x}+\tau_{j}^{y}\tau_{j+1}^{y}+\frac{1}{2}\tau_{j}^{z}\tau_{j+1}^{z}] \nonumber \\
&-&h_{0}^{eff}\sum_{j=1}^{N/2}\tau_{j}^{z}-h_{1}^{eff}\sum_{j=1}^{N/2}(-1)^{j}\tau_{j}^{z}, \label{effective-hamiltoni}
\end{eqnarray}
where $h_{0}^{eff}=h-J_{AF}+\frac{J_{F}}{4}$ and
$h_{1}^{eff}=\delta J_{AF}$. Thus, the effective Hamiltonian is
nothing but the XXZ Heisenberg chain, with anisotropy $\Delta=1/2$
in a uniform and staggered longitudinal magnetic field. The full
phase diagram of this model has been investigated by F.
C. Alcaraz and A. L. Malvezzi \cite{Alca 95} and the nature of the
ground state phase transition has been pointed in
Ref. \cite{Okam 96}. The other properties also could be
found in recent works such as Refs. \cite{oshikawa97,
Japaridze09, fouet04, mahdavifar07}. To find clearer picture
for the low energy spectrum of the effective model
(\ref{effective-hamiltoni}), using the numerical Lanczos method we
have computed the energy gap of the effective model. The numerical
results, have been shown in Fig.\ref{energy-gap-effective} for a
chain of lengths $N=16$, $20$ and the exchange parameters
$J_{F}=1.0$, $J_{AF}=9/2$, and $\delta=1/9$. As it is seen from
Fig.\ref{energy-gap-effective}, at $h_0^{eff}=0$ or
$h=J_{AF}-\frac{J_{F}}{4}$, the effective model reduces to the XXZ
spin chain in a staggered longitudinal magnetic field. The ground
state of this model is in the gapped N\'{e}el phase. By
changing $h_0^{eff}$ from zero i.e decreasing (increasing) $h$
from the value $h=J_{AF}-\frac{J_{F}}{4}$, the energy gap
decreases and reaches to zero at finite values of the magnetic
field $h_0^{eff, -}$ ($h_0^{eff, +}$). Consequently, there is a
plateau in the magnetization curve of the effective model at
$0<h_0^{eff}<h_0^{eff, +}$. By more decreasing (increasing) of
$h$, the energy gap remains zero dawn(up) to critical field
$h_0^{eff, 1}$ ($h_0^{eff, 2}$) and, after these gapless regions
it is re-opened and behaves linearly with the magnetic field
$h_0^{eff}$. Correspondingly, in these regions the magnetization
of the effective model starts to change from zero and is saturated
at the critical field ($h_0^{eff, 2}$).


\section{Field theory predictions}

In this section we will study the model (\ref{hamiltoni}) in the
language of two continuum field theories. We have employed the
nonlinear $\sigma$ model and Bosonization approaches to obtain the energy gap, the
critical fields and magnetization plateaus of the model.

\subsection{Nonlinear $\sigma$ model}

The $O(3)$ nonlinear $\sigma$ model (NL$\sigma$M) is a
semiclassical approach which is based on the certain properties of
a field theory in (1+1) dimension. In 1983 Haldane predicted
theoretically the existence of a finite gap between the ground
state and the first excited state of antiferromagnetic Heisenberg
integer spin chains \cite{haldane83}. He also conjectured that the
half-odd integer spin chains are gapless. The Hamiltonian of a
homogeneous spin chain is mapped to a $O(3)$ NL$\sigma$M with an
additional topological term. The topological term is $0$ if spins
of the chain are integer and $\pi$ if they are half-odd integer.

Following we will investigate the low temperature behaviors of the
alternating spin chain (\ref{hamiltoni}). For convenient on the
incoming calculations let us write the Hamiltonian (\ref{hamiltoni})
as the following form
\begin{eqnarray}
\nonumber &&H = -J_F\sum_{j=1}^{N/4}({\bold S}_{2j-1}^{(1)}\cdot {\bold S}_{2j-1}^{(2)}+{\bold S}_{2j}^{(1)}\cdot {\bold S}_{2j}^{(2)})\\
\nonumber&&+J_{AF}\sum_{j}^{N/4}[(1+\delta){\bold S}_{2j-1}^{(2)}\cdot {\bold S}_{2j}^{(1)}+(1-\delta){\bold S}_{2j}^{(2)}\cdot {\bold S}_{2j+1}^{(1)}]\\
&&+h\sum_j{\bold S}_{j}^{z}\label{H-1}
\end{eqnarray}
where $0\leq\delta\leq1$. Using the spin coherent states
representation for the spin operators, i.e ${\bf S}=s
{\mathbf\Omega}$ we can write the Hamiltonian in terms of the
classical spin vectors. By introducing three classical fields ${\mathbf n}$, ${\mathbf L}$, and ${\mathbf\Delta}$
the classical vectors are written as the following forms:
\begin{eqnarray}
\nonumber\mathbf{\Omega}_{2j-1}^{(i)}&=&-\mathbf{n}_{2j-1}(1-|\frac{a}{s}\mathbf{L}_{2j-1}|^2)^{1/2}+\frac{a}{s}\mathbf{L}_{2j-1}\\
\nonumber&&-(-1)^i\frac{a}{s}\mathbf{\Delta}_{2j-1},\\
\nonumber\mathbf{\Omega}_{2j}^{(i)}&=&\mathbf{n}_{2j}(1-|\frac{a}{s}\mathbf{L}_{2j}|^2)^{1/2}+\frac{a}{s}\mathbf{L}_{2j}+(-1)^i\frac{a}{s}\mathbf{\Delta}_{2j},\\
\label{mapping}
\end{eqnarray}
where $i=1,2$ and $a$ is the lattice constant. Typically, for a quantum antiferromagnetic Heisenberg chain
it is convenient to write the spin vectors in terms of a unimodular N\'{e}el field
(${\mathbf n}$) and one ferromagnetic canting field (${\mathbf
L}$). However for an alternating AF-F
Heisenberg spin chain, the main difficulty arises from the fact
that it is inhomogeneous and writing a continuous action out of the
discrete Hamiltonian is not a trivial task \cite{Bosq00}. Using
another field which describes the variation of the N\'{e}el
fields inside the two spin-blocks such as ${\mathbf\Delta}$ we can
map the Hamiltonian (\ref{H-1}) to the following $O(3)$
NL$\sigma$M:
\begin{equation}
{\cal A}=\frac{1}{2g}\int_{0}^L\int_0^{L_T} dxdx_0[(\partial_x \mathbf{n})^2+(\partial_{x_0} \mathbf{n})^2]-i\Theta W,
\end{equation}
where $W$ is the winding number, $L_T=c\beta$, $x_0=c\tau$, and we have considered the case $h=0$. By defining $\alpha=\frac{J_F}{J_{AF}}$, the coupling constant $g$, and the velocity of spin excitations $c$ are given in terms of $\alpha$ and $\delta$ as:
\begin{eqnarray}
\nonumber g&=&\frac{1-\frac{\delta^2}{2(1+\alpha)}}{s\left(1-\frac{\delta^2}{4}-\frac{1}{2(1+\alpha)}\right)^{1/2}},\\
c&=&J_{AF} a s\left(1-\frac{\delta^2}{4}-\frac{1}{2(1+\alpha)}\right)^{1/2}.
\end{eqnarray}

The topological $\Theta$ term is obtained as follow:
\begin{equation}
\Theta=2\pi
\bigg(\frac{s\delta\alpha}{1+\alpha-\frac{\delta^2}{2}}\bigg),
\end{equation}
where . For $\alpha=0$ or $\delta=0$
the topological $\Theta$ term is zero and the model is always
gapped and the spin excitations velocity is $\frac{\sqrt{2}}{2}J_{AF} a s$.  Moreover, the $\alpha=1$ and $\delta=0$ is the special
case of AF-F Heisenberg spin chains which the topological term is
always $0$. This result is in good agreement with
the result presented in Ref. \cite{Bosq00}. For any other values of $\alpha$ and
$\delta$, the Hamiltonian is mapped to a non-integrable $O(3)$
NL$\sigma$M with $\Theta$ in the interval $[0,\pi]$. This model is
also gapped and the gap value is dependent on $\alpha$ and
$\delta$.

Now, let us study the model in presence of the magnetic field.
Since, the model is
always gapped in the absence of a magnetic field we are allowed
to consider $\Theta=0$.
The Hamiltonian
(\ref{H-1}) is mapped to the following NL$\sigma$M which the
effect of the magnetic field is clearly seen:
\begin{equation}
{\cal A}=\frac{1}{2g}\int d^2x[(\partial_x
\mathbf{n})^2+(\partial_{x_0}
\mathbf{n}-\frac{i}{c}~\bf{h}\times\mathbf{n})^2], \label{field-A}
\end{equation}

To get more physical insight from the effects of the magnetic
field, it may be more physically transparent to work with two
fields ($\theta$, $\phi$) where $\mathbf{n}=(\sin\theta
\sin\phi,\sin\theta \cos\phi, \cos\theta)$. Here $\theta$ is
co-latitude and $\phi$ is azimuthal angles. Selecting magnetic
field in $z$ direction, the action (\ref{field-A}) takes the
following form;
\begin{equation}
{\cal A}=\frac{1}{2g}\int
d^2x\{(\partial_{\mu}\theta)^2+\sin^2\theta\left[(\phi^{\prime})^2+\frac{1}{c^2}(\phi^{\cdot}-i
h)^2\right]\}, \label{action-theta}
\end{equation}
where $\theta$ is the angle of $\mathbf{h}$ and $\mathbf{n}$. In
the term, $\mathbf{h}\times\mathbf{n}$, the magnetic field induces
a hard-axis anisotropy. In other word, the magnetic field try to
align all spins with $\mathbf{n}$ in the plane normal to $h$. Thus
for high field regime, $h>|\phi^{\cdot}|$, the deviation of
N\'{e}el field $\mathbf{n}$ from the plane is small. Thus an
expansion to quadratic order in $\vartheta$ is valid. The action
is
\begin{eqnarray}
\nonumber&&{\cal A}=\frac{1}{2g}\int_0^{L}\int_0^{L_T}d^2\mu\bigg(-\bar{h}^2-\phi \partial_{\mu}^2
\phi-2 i \bar{h}\partial_{x_0}\phi\\&&
+(\partial_{\mu}\vartheta)^2-\vartheta^2
\left[(\partial_{x_0}\phi)^2-\bar{h}^2-2 i \bar{h}\partial_{x_0}\phi\right]\bigg),
\end{eqnarray}
where ${\bar h}=h/c$ and
the fluctuations have been separated to the in-plane and out-of-plane fluctuations.

Using a spin stiffness analysis, $1/N$ expansion and a
renormalization group approach, it has been already computed the
magnetization and spin correlation functions of a spin ladder in
an applied magnetic field\cite{Loss 95, Norm}. The magnetization
($\sum_i\frac{\langle S_i\rangle}{N}$) is given by
$M=\frac{1}{N}\frac{\partial F}{\partial h}$, where
$F=-\frac{1}{\beta}\ln Z$ is the Helmholtz free energy and
$Z=\int{\cal D}\vartheta{\cal D}\phi e^{-{\cal A}}$. The separable
nature of fluctuations allows us to give the results of
$M=M_o+M_i$ which is a summation of both out of plane and in plane
contributions. At low enough temperatures and small value of $c/h$
one can find that the out of plane contribution is a constant,
$M_o=1/2$, which is correspond to a uniform state. The in plane
contribution has two terms,  one is linear in the field and the
other has a sawtooth form (See
Ref. \cite{Norm}). Totally, $M_i$ is a step-like form which
the width of the steps scales as $1/N$. Thus the sawtooth form is
the finite size corrections and, in the thermodynamic limit
$N\rightarrow\infty$ the in plane magnetization is only linear.

For our tetramerized spin chain the magnetization is given by
\begin{equation}
M\simeq-1+\frac{h}{g c}.
\end{equation}
Application of an enough high magnetic field will cause the spin alignment, or saturation with a maximum magnetization, $M_s=s$. This effect should be considered in the NL$\sigma$M approach by a lagrange multiplier.
At zero magnetic field the system is gapped and the magnetization
is zero. Tuning magnetic field decreases the gap till first
critical field, $h_{c_1}$ where magnetization start to increase.
At the second critical field, where the magnetization saturated,
the gap reopens and spins are fully aligned in the magnetic field
direction.

The critical fields which are attained by means of NL$\sigma$M are as follows:
\begin{eqnarray}
\nonumber h_{c_1}&=&gc,\\
h_{c_2}&=&\frac{3}{2}gc.
\end{eqnarray}


\subsection{Bosonizaion}

In this section we concentrate our attentions to the low-energy
and long wavelength excitations by using bosonization language.
Let us consider the Hamiltonian (\ref{effective-hamiltoni}). In
our analysis of the model (\ref{effective-hamiltoni}) we closely
follow the route developed in the Ref. \cite{Japaridze09}.
In the absence of both magnetic fields, $h_0^{eff}=0$ and
$h_1^{eff}=0$ we have the XXZ spin-1/2 chain with ferromagnetic
coupling and anisotropy parameter $\Delta=1/2$. This model is
critical and the long-wavelength excitations are described by the
following Gaussian theory;
\begin{equation}
{\cal H}=\frac{v}{2}\int dx [(\partial_x \theta)^2+(\partial_x \phi)^2],
\end{equation}
where $\theta(x)$ and $\phi(x)$ are dual bosonic fields, $\partial_t\phi=v\partial_x\theta$, and satisfying the following commutation relations
\begin{eqnarray}
&&[\phi(x),\theta(y)]=i\Theta(y-x),\\
&&[\phi(x),\theta(x)]=\frac{i}{2}.
\end{eqnarray}
$v$ is the spin excitation's velocity and fixed by the Bethe ansatz solution as
\begin{equation}
v=(\frac{J_F}{2})\frac{K}{2K-1}\sin(\frac{\pi}{2K}),
\end{equation}
where L\"{u}ttinger {\it parameter} $K$ (inverse of Bosons
radius \cite{Egge 92}) is given as a function of $\Delta$ as
\begin{equation}
K=\frac{\pi}{2\arccos\Delta}.
\end{equation}
Thus $K$ is increased monotonically along the XXZ critical line
$-1<\Delta\leq1$ from its minimal value $K=1/2$ (for the isotropic
antiferromagnetic $\Delta=-1$ case) to unity at $\Delta=0$ (for the
noninteracting case) and goas to infinity at $\Delta=1$ which is
the ferromagnetic instability point. Meanwhile the boson radius
has its maximum value $(2\pi)^{-1/2}$ for the isotropic
antiferromagnetic chains and decreases by changing in $\Delta$ and
goes to zero at the ferromagnetic instability point.

The maximum value of the spin excitation velocity occurs where
$K\sim1/2$ or $\Delta=-1$. It decreases monotonically by
increasing $K$ and reaches to the zero value at $K=\infty$
($\Delta=1$).

The continuum limit of the Hamiltonian (\ref{effective-hamiltoni})
is obtained by writing spin operators in terms of bosonic fields:
\begin{eqnarray}
\nonumber&&\tau_n^z=\sqrt{\frac{K}{\pi}}\partial_x\phi+\frac{(-1)^n~a}{\pi}\sin[\sqrt{4\pi K}\phi(x)],\\
\nonumber&&\tau_n^+=\frac{1}{\sqrt{2\pi}}e^{-i\sqrt{\frac{\pi}{K}}\theta(x)}((-1)^n~b\sin[\sqrt{4\pi K}\phi(x)]+c),\\
\label{bosonization}
\end{eqnarray}
where $a$, $b$ and $c$ are non-universal real constant in the
order of unity and depends on the parameter $\Delta$ \cite{Hiki
98}. In above transformations we have made the rotation
$\tau_n^{\pm}\rightarrow (-1)^n\tau_n^{\pm}$ and
$\tau_n^z\rightarrow\tau_n^z$ on the standard bosonic version of
the spin operators which is found extensively in the literature.
Using (\ref{bosonization}), the effective Hamiltonian density is
given as:
\begin{eqnarray}
\nonumber{\cal H}&=&\frac{v}{2}[(\partial_x\theta)^2+(\partial_x\phi)^2]\\
&&+\frac{h_1^{eff}}{\pi a_0}\sin[\sqrt{4\pi K}\phi]-h_0^{eff}\sqrt{\frac{K}{\pi}}\partial_x\phi.
\end{eqnarray}

The mapped model is nothing but the massive sin-Gordon model with
an additional topological term. Let us first consider the
sin-Gordon model without the topological term $h_0^{eff}=0$ or at
the magnetic field value $h=J_{AF}-\frac{J_F}{4}$. As it has been
discussed in many places, in the interval $1<K<2$ the spectrum of
the SG model contains soliton and anti-soliton with mass ${\cal
M}$. The exact relation between the soliton mass and the bare mass
$h_1^{eff}$ is as follows \cite{Zamo 95}
\begin{eqnarray}
\label{Smass}&&{\cal M}={\cal C}(K)h_1^{{eff}^{\frac{1}{2-K}}},\\
\nonumber&&{\cal C}(K)=\frac{2~\Gamma(\frac{1}{2\nu})}{\sqrt{\pi}\Gamma(\frac{1}{2}+\frac{1}{2\nu})}
[\frac{\Gamma(1-\frac{K}{2})}{2~\Gamma(\frac{K}{2})}]^{\frac{1}{2-K}}.
\end{eqnarray}
By substituting $K=3/2$ and $h_1^{eff}=\delta J_{AF}$ in
(\ref{Smass}) the soliton mass or equivalently the excitation gap
is given as ${\cal M}\sim\frac{2}{J_F}(\delta J_{AF})^2$. Turning
$h_0^{eff}$ add a gradient term to the gapped SG model. This term
creates a shift on the field $\phi$. Actually, in the absence of
the topological term the excitation spectrum of the model is
gapped and the field $\phi$ is sticken on one of the minima where
$\langle\sin\sqrt{6\pi}\rangle=-1$. This one is correspond to a
staggered AF order in the effective spin chain. Tuning the uniform
field $h_0^{eff}$ the number of particle is decreased and the
minima is shifted. Challenging between two uniform and staggered
field causes the system experiences a different phase when
$|h_0^{eff}|>{\cal M}$ \cite{Japa 79}. This phase
transition is occurred at $h_c^-$ where the excitation gap
reopened.
\begin{figure}
\centerline{\includegraphics[width=8cm,angle=0]{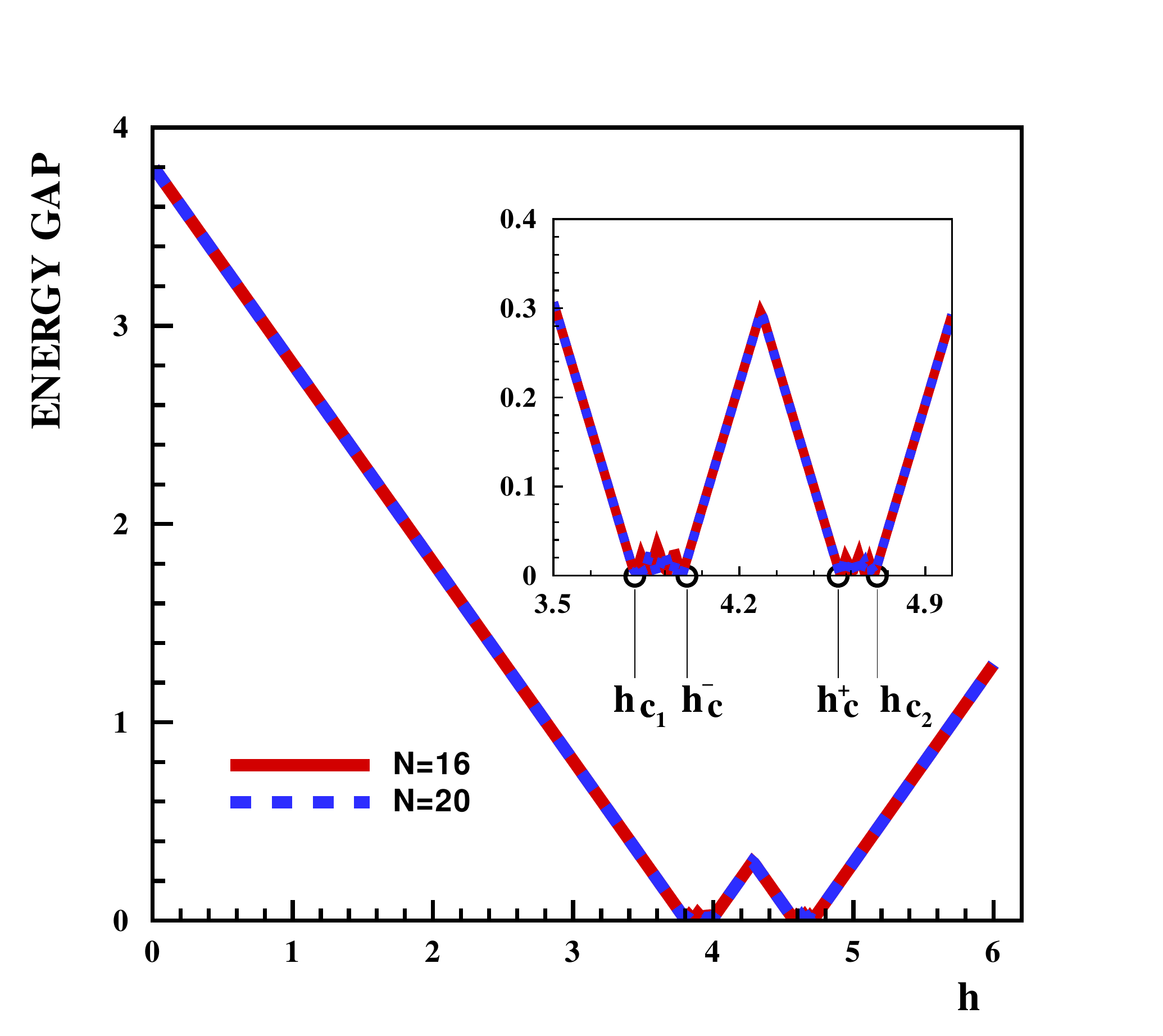}}
\caption{(Color online.) Difference between the two lowest energy levels of the original Hamiltonian versus the magnetic field $h$, for chains
with different lengths $N=16,20$ and exchange parameters $J_{F}=1.0$, $J_{AF}=9/2$ and $\delta=1/9$.}\label{energy-gap}
\end{figure}

Consequently the width of the magnetization plateau is obtained as
\begin{equation}
h_{c}^{+}-h_{c}^{-}\simeq \frac{4}{J_F}(\delta J_{AF})^2,
\end{equation}
where $h_c^+$ and $h_c^-$ are the middle critical field and are
given
\begin{equation}
h_c^{\pm}=J_{AF}(1+\frac{\alpha}{4}\pm\frac{2\delta^2}{\alpha}).
\end{equation}

Summarizing, implementing two continuum filed theories we found that the model has four critical fields.


\section{Numerical results}

To explore the nature of the spectrum and the
quantum phase transition, we have used Lanczos method to diagonalize
numerically chains with length up to $N=28$.

First, we have computed the three lowest energy eigenvalues of
chains with $J_{F}=1.0$, different values of the length and
antiferromagnetic exchanges. To get the energies of the few lowest
eigenstates we consider chains with periodic boundary conditions.

\begin{figure}
\centerline{\includegraphics[width=8cm,angle=0]{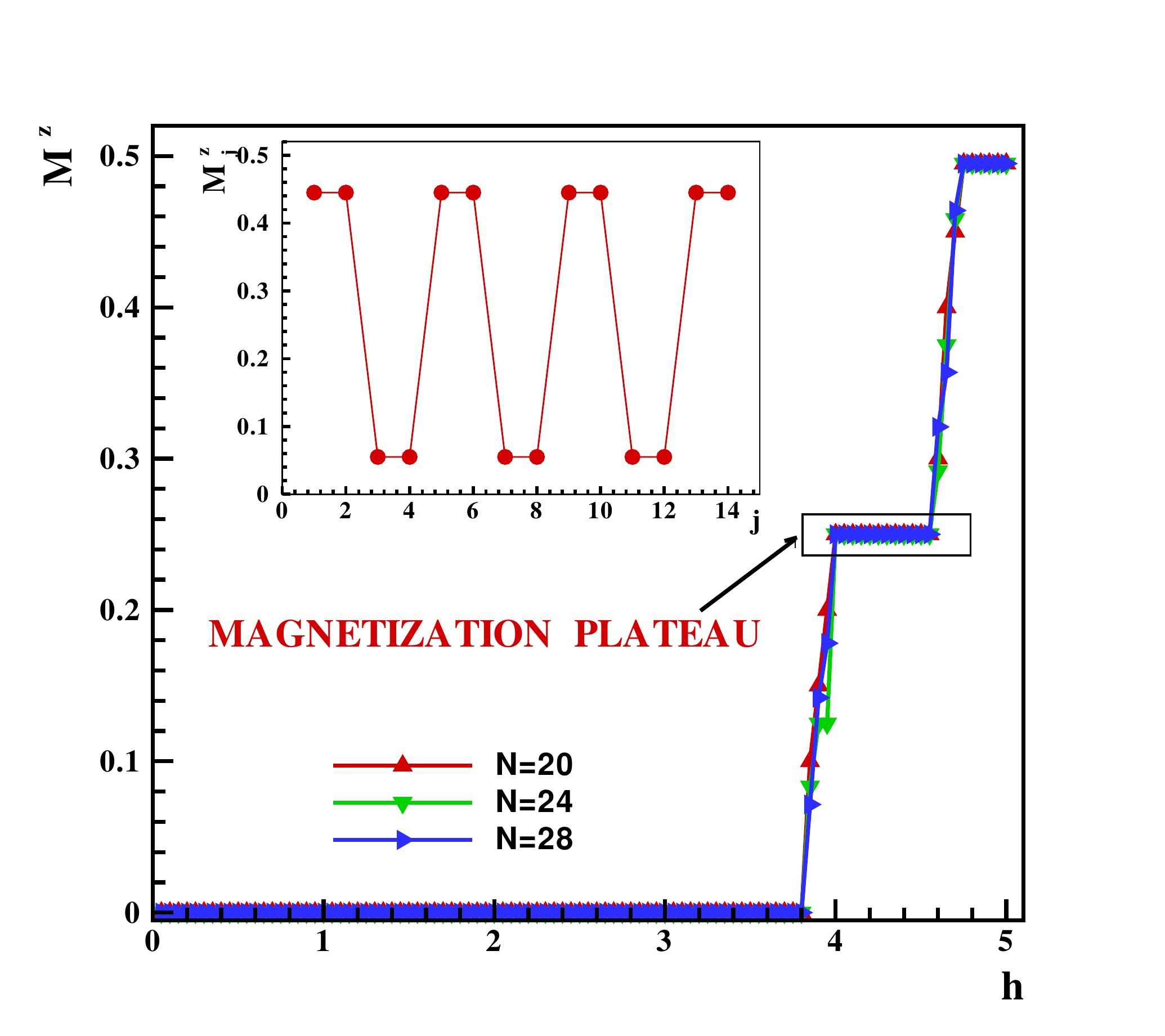}}
\caption{(Color online.) The magnetization along the field, $M^z$, as a
function of the applied magnetic field $h$, for chains with
exchanges $J_{F}=1.0$ , $J_{AF}=9/2$, $\delta=1/9$ and lengths
$N=20, 24, 28$. In the inset, the magnetization on site as a
function of the site number $j$ is plotted for a value of the
magnetic field in the region of the to mid-plateau. }
\label{Magnetization}
\end{figure}

In Fig.\ref{energy-gap}, we present results of these calculations
for the exchanges $J_{F}=1.0, J_{AF}=9/2, \delta=1/9$ and chain
sizes $N=16,20$. We define the excitation gap as a gap on the
first excited state. As it is seen in Fig. \ref{energy-gap}, in
strong limit of antiferromagnetic exchange, this difference is
characterized by the indistinguishable (within the used numerical
accuracy) dependence on the chain length and shows an universal
linear decrease with increasing magnetic field. At $h=0$ the
spectrum of the model is gapped. Turning the magnetic field the
energy gap decreases linearly with $h$ and vanishes at
$h_{c_{1}}$. This is the first level crossing between the
ground-state energy and the first excited state energy. The
spectrum remains gapless for $h_{c_{1}}<h<h_{c}^{-}$, whereas the
gap is reopened when $h>h_{c}^{-}$. After an increasing and a
decreasing the spin gap goes to zero and vanishes at
$h_{c}^{+}$. More increasing in the field $h>h_{c}^{+}$, the
spectrum remains gapless up to the critical saturation field
$h_{c_{2}}$. Finally, at $h>h_{c_{2}}$ the gap is reopened and for
a sufficiently large field becomes proportional to $h$.
Oscillations of the energy gap in regions $h_{c_{1}}<h<h_{c}^{-}$
and $h_{c_{2}}<h<h_{c}^{+}$ are the result of level crossings in
finite size systems. To find the critical fields we have used the
phenomenological renormalization group (PRG)
method \cite{mahdavifar08}. The critical field values are given as follows:
\begin{eqnarray}
h_{c_{1}}&=&3.80\pm 0.01,~~~~~~~~~~~h_{c_{2}}=4.71\pm 0.01,\nonumber  \\
h_{c^{-}}&=&3.99\pm 0.01,~~~~~~~~~~~h_{c^{+}}=4.58\pm
0.01.\label{critical-fields}
\end{eqnarray}
To study the magnetic order of the ground state of the system, we
start with the magnetization process. First, we have implemented
the Lanczos algorithm on finite chains to calculate the lowest
eigenstate. The magnetization along the field axis is defined as
\begin{eqnarray}
M^{z}=\frac{1}{N}\sum_{j=1}^{N}\langle GS|S_{j}^{z}|GS \rangle,
\label{magnetization}
\end{eqnarray}
where the notation $\langle Gs|...|Gs\rangle$ represents the
ground state expectation value. In Fig.~\ref{Magnetization}, we
have plotted $M^{z}$ as a function of the magnetic field $h$, and
for a chain with exchange parameters $J_{F}=1.0, J_{AF}=9/2,
\delta=1/9$ and different lengths $N=20, 24, 28$ . As it is
clearly seen in Fig.~\ref{Magnetization} besides the standard
singlet and saturation plateaus at $h<h_{c_{1}}$ and $h>h_{c_{2}}$
respectively, we observe a plateau at $M=\frac{1}{2}M_{sat}$.
Observed oscillations of the magnetization at $h_{c_{1}}< h
<h_{c}^{-}$ and $h_{c}^{+} < h< h_{c_{2}}$ result from the level
crossing between the ground and the first excited states of this
model in the gapless phases. To check that the mid-plateau is not
a finite size effect, we performed the size
scaling \cite{zhitomirsky00} of its width and found that the size
of the plateau interpolates to finite value when
$N\longrightarrow\infty$. In the inset of Fig.~\ref{Magnetization}
we have plotted the magnetization on site, $M_{j}^{z}=\langle
GS|S_{j}^{z}|GS \rangle$, as a function of the site number "$j$"
for a value of the magnetic field corresponding to the plateau at
$M^{z}=0.5M_{sat}^{z}$. To obtain an accurate estimate of the
function $M_{j}^{z}$, we have calculated it for system sizes of
$N=12, 16, 20, 24, 28$. The thermodynamic limit ($N \rightarrow
\infty$) of the finite size results are obtained by extrapolation
method and used for plotting. As we observe the system shows a
well pronounced modulation of the on site magnetization, where
magnetization on odd bonds  is larger than on even bonds. This
distribution remains almost unchanged within the plateau for
$h_{c}^{-} < h< h_{c_{+}}$.

By analyzing the numerical results on the energy gap
(Fig.~\ref{energy-gap}), we found that the spectrum is gapped in
absence of the magnetic field which is one of the properties of
the Haldane phase. The Haldane phase can be recognized from
studying the string order parameter. The string correlation
function in a chain of length $N$, defined only for odd $l$,
is\cite{hida99}
\begin{eqnarray}
O_{Str}(l, N)=-\langle exp\{i\pi \sum_{2j+1}^{2j+l+1} S_{k}^{z}\}\rangle.
\label{string}
\end{eqnarray}
\begin{figure}
\centerline{\includegraphics[width=8cm,angle=0]{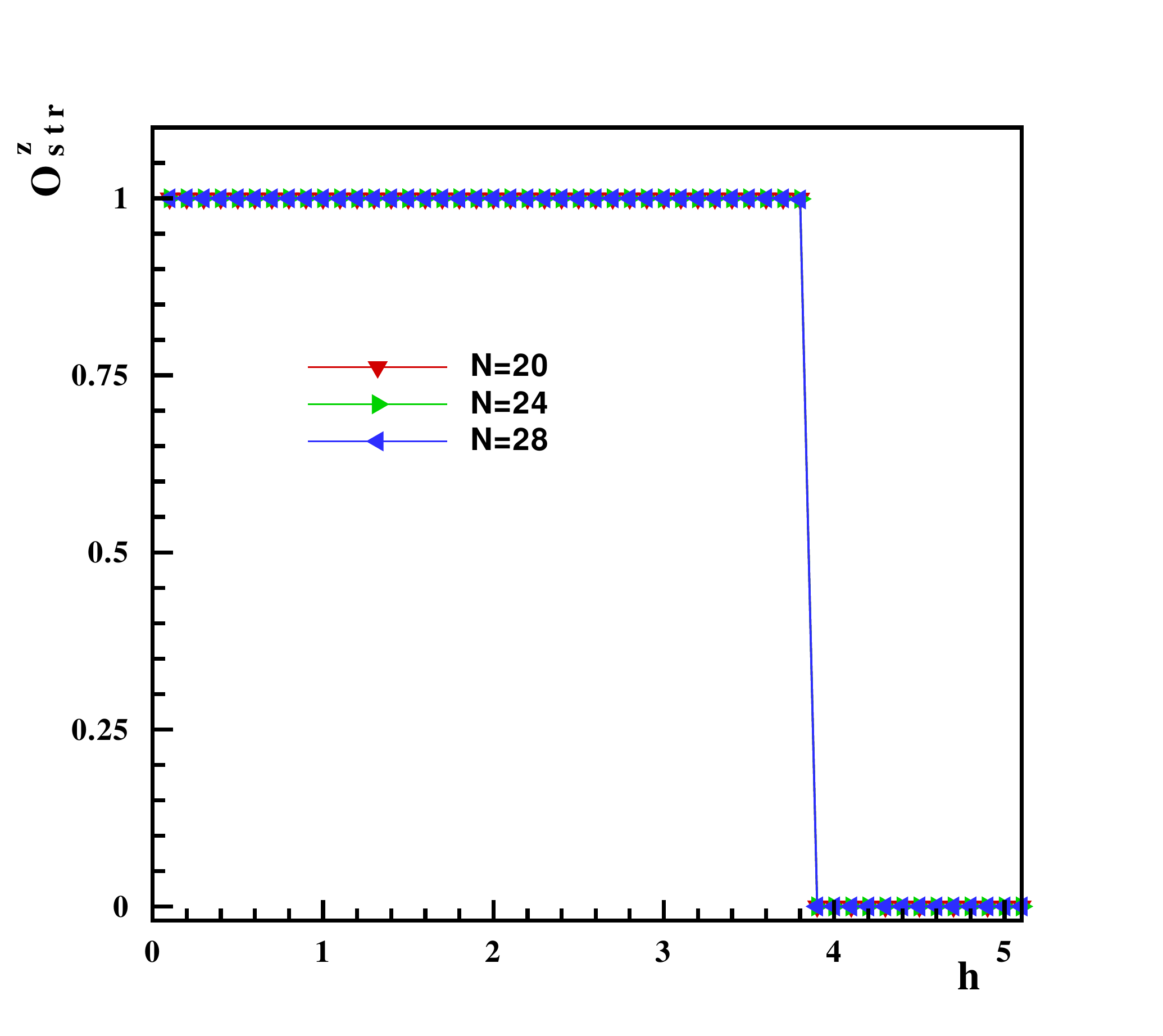}} \caption{(Color
online.) The string correlation function, $O_{Str}(l, N)$, as a
function of the applied magnetic field $h$, for chains with
exchanges $J_{F}=1.0$ , $J_{AF}=9/2$, $\delta=1/9$ and lengths
$N=20, 24, 28$.  } \label{String}
\end{figure}
In particular, we calculated the string correlation function for
different finite chain lengths. Since the present model has a
$SU(2)$ symmetry in the absence of a magnetic field, we only
consider the $z$ component of the string correlation function.  In
Fig.~(\ref{String}),
 we have plotted $O_{Str}(l, N)$ as a function of $h$ for the chain with exchanges $J_{F}=1.0$
, $J_{AF}=9/2$, $\delta=1/9$ and lengths $N=20, 24, 28$. As can be
seen from this figure, at $h<h_{c_{1}}$, the string correlation
function $O_{Str}(l, N)$ is saturated and  the tetrameric chain
system is in the Haldane phase. The Haldane phase remains stable
even in the presence of a magnetic field less than $h_{c_{1}}$.

\begin{figure}
\centerline{\includegraphics[width=8cm,angle=0]{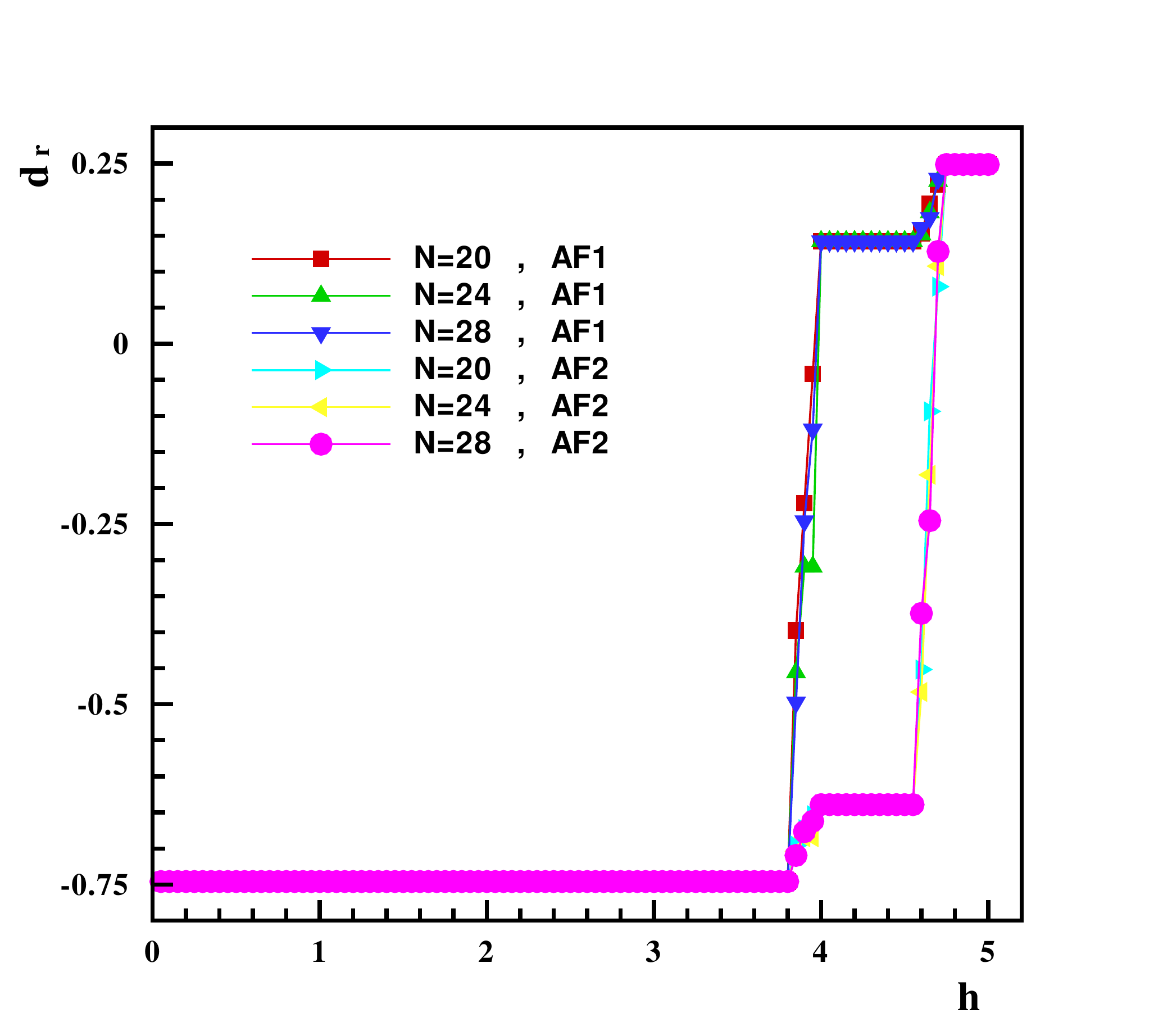}}
\caption{(Color online.) The AF-bond dimerization order parameter
as a function of the applied field $h$, for chains with  exchanges
$J_{F}=1.0$ , $J_{AF}=9/2$, $\delta=1/9$ and lengths $N=20, 24,
28$. } \label{dimerization}
\end{figure}

An additional insight into the nature of different phases can be
obtained by studying the correlation functions. We define the following weak and strong bond
dimerization order parameters;
\begin{equation}
d^{w}_r=\frac{4}{N}\sum_{j_{odd}=1,3,5,\dots}^{N/2}\langle
GS|{\bf S}_{2j-1}\cdot{\bf S}_{2j}|GS \rangle , \label{dimerization_O}
\end{equation}
and
\begin{equation}
d^{s}_{r}=\frac{4}{N}\sum_{j_{even}=2,4,6,\dots}^{N/2}\langle
GS|{\bf S}_{2j-1}\cdot{\bf S}_{2j}|GS \rangle, \label{dimerization_E}
\end{equation}
where summations are taken over the weak and strong
antiferromagnetic-bonds. In Fig.~(\ref{dimerization}) we have
plotted $d^{w}_r$ and $d^{s}_r$ versus magnetic field $h$ for
chain of lengths $N=20, 24, 28$ with the exchange parameters
$J_{F}=1.0, J_{AF}=9/2$ and $\delta=1/9$. As it is seen from this
figure, at $h<h_{c_{1}}$ spins on all antiferromagnetic-bonds are
in a singlet state $d_{r}^{w}=d_{r}^{s} \simeq -0.75$, while at
$h>h_{c_{2}}$, $d_{r}$ is equal to the saturation value
$d_{r}^{w}=d_{r}^{s}\sim 1/4$ and the ferromagnetic long-range
order along the field axis is present.
However, in the considered case of strong antiferromagnetic
exchanges ($J_{AF}^{\pm}\gg J_{F}$) and high critical fields
quantum fluctuations are substantially suppressed and calculated
averages of on-antiferromagnetic-bond spin correlations are very
close to their nominal values.

On the other hand, for intermediate values of the magnetic field,
at $h_{c_{1}}<h<h_{c_{2}}$ the data presented in
Fig.~(\ref{dimerization}) gives us a possibility to trace the
mechanism of singlet-pair melting with increasing magnetic field.
As it follows from Fig.~(\ref{dimerization}) at $h$ slightly above
$h_{c_{1}}$ spin singlets pairs start to melt in all
antiferromagnetic bonds simultaneously and almost with the same
intensity. With further increase of $h$ melting of weak
antiferromagnetic-bonds gets more intensive, however at
$h=h^{-}_{c}$ the process of melting stops. As it is seen in
Fig.~(\ref{dimerization}) weak antiferromagnetic-bonds are
polarized, however their polarization is far from the saturation
value $d^{w}_r\simeq 0.15$, while the strong
antiferromagnetic-bonds still manifest strong on-site singlet
features with $d^{s}_r \simeq -0.65$.  Moratorium on melting stops
at $h=h^{+}_{c}$, however for $h>h^{+}_{c}$ strong
antiferromagnetic-bonds start to melt more intensively while the
polarization of weak antiferromagnetic-bonds increases slowly.
Finally at $h=h_{c_{2}}$ both subsystems of
antiferromagnetic-bonds achieve an identical, almost fully
polarized state. Note, that the almost symmetric fluctuations in
on-antiferromagnetic-bonds correlations, increase in $d^{w}_r$ at
$h \leq h^{-}_{c}$ decrease in $d^{s}_r$ at $h \geq h^{+}_{c}$
reflect the enhanced role of quantum fluctuations in vicinity of
quantum critical points.

\begin{figure}
\centerline{\includegraphics[width=8cm,angle=0]{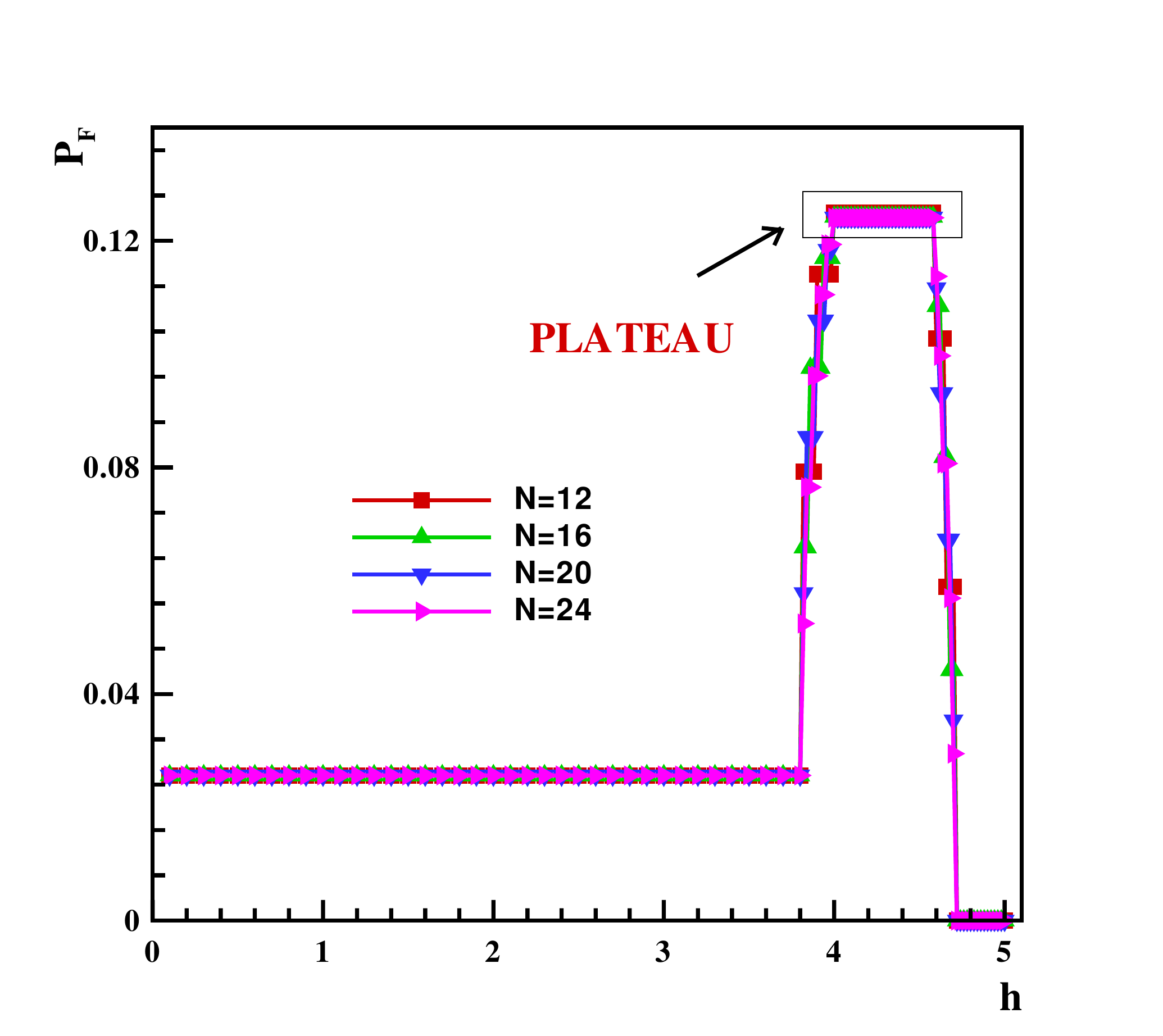}}
\caption{(Color online.) The F-dimer order parameter as a
function of the applied field $h$, for the chains with  exchanges
$J_{F}=1.0$ , $J_{AF}=9/2$, and $\delta=1/9$ and lengths $N=12$, $16$, $20$, and $24$. The appearance of a non-zero plateau in the curve is clear.} \label{F-dimer}
\end{figure}

In our previous paper \cite{abouie08}, we introduced a
mean field order parameter which can distinguish a gapless
LL phase from the other gapped phases. This order parameter is the F-dimer order parameter which is
defined as
\begin{eqnarray}
P_F=Re\la S^{-}_{2n}S^{+}_{2n+1}\ra  \label{ord.pa}.
\end{eqnarray}
The F-dimer order parameter has a considerable value in the
L\"{u}ttinger liquid phase and behaves differently in the other
gapped phases.  The effects of a small value of space modulation
on this parameter has been shown in Fig.~(\ref{F-dimer}). We have
plotted $P_F$ versus magnetic field $h$ for chain of lengths
$N=12$, $16$, $20$, and $24$ with the exchange parameters
$J_{F}=1.0$, $J_{AF}=9/2$, and $\delta=1/9$. As it is seen, in the
Haldane phase, $h<h_{c_{1}}$, the quantum fluctuations suppress
the ferromagnetic correlations and the F-dimer parameter is close
to zero-value. Right after the first critical field, the F-dimer
parameter increases rapidly up to the second critical field. In
the intermediate region, $h_{c^{-}}<h<h_{c^{+}}$, the F-dimer
parameter shows a non-zero plateau which behavior is the same as
the other parameters. By more increasing field, the F-dimer
decreases and goes to zero at the saturation critical field
$h=h_{c_{2}}$. The intermediate region is a gapped phase and we
expect the zero-value for the defined LL parameter, $P_F$. However
the gap does not affect on the behavior of $P_F$ and a plateau
appears in the curve.

\section{Pair-wise entanglement}

In this section we focus on the entanglement of two spins in
different phases of the system. The entanglement which has no a
classical counterpart is employed to study the quantum
correlations of different states. Concurrence is a measure of the
bipartite entanglement which is defined as
following \cite{Wooters98,Amico08}
\begin{eqnarray}
C_{lm}&=&2~max\{0, C_{lm}^{(1)}, C_{lm}^{(2)}\},
\label{concurrence}
\end{eqnarray}
where
\begin{eqnarray}
C_{lm}^{(1)}&=& \sqrt{(g_{lm}^{xx}-g_{lm}^{yy})^{2}+(g_{lm}^{xy}+g_{lm}^{yx})^{2}} \nonumber \\
&-&\sqrt{(\frac{1}{4}-g_{lm}^{zz})^{2}-(\frac{M_{l}^{z}-M_{m}^{z}}{2})^{2}},\nonumber \\
C_{lm}^{(2)}&=& \sqrt{(g_{lm}^{xx}+g_{lm}^{yy})^{2}+(g_{lm}^{xy}-g_{lm}^{yx})^{2}} \nonumber \\
&-&\sqrt{(\frac{1}{4}+g_{lm}^{zz})^{2}-(\frac{M_{l}^{z}+M_{m}^{z}}{2})^{2}}
\label{concurrence}
\end{eqnarray}
and $g_{lm}^{\alpha\beta}=\langle S_{l}^{\alpha}
S_{m}^{\beta}\rangle$ is the correlation function between spins $l$ and $m$.
\begin{figure}
\centerline{\includegraphics[width=8cm,angle=0]{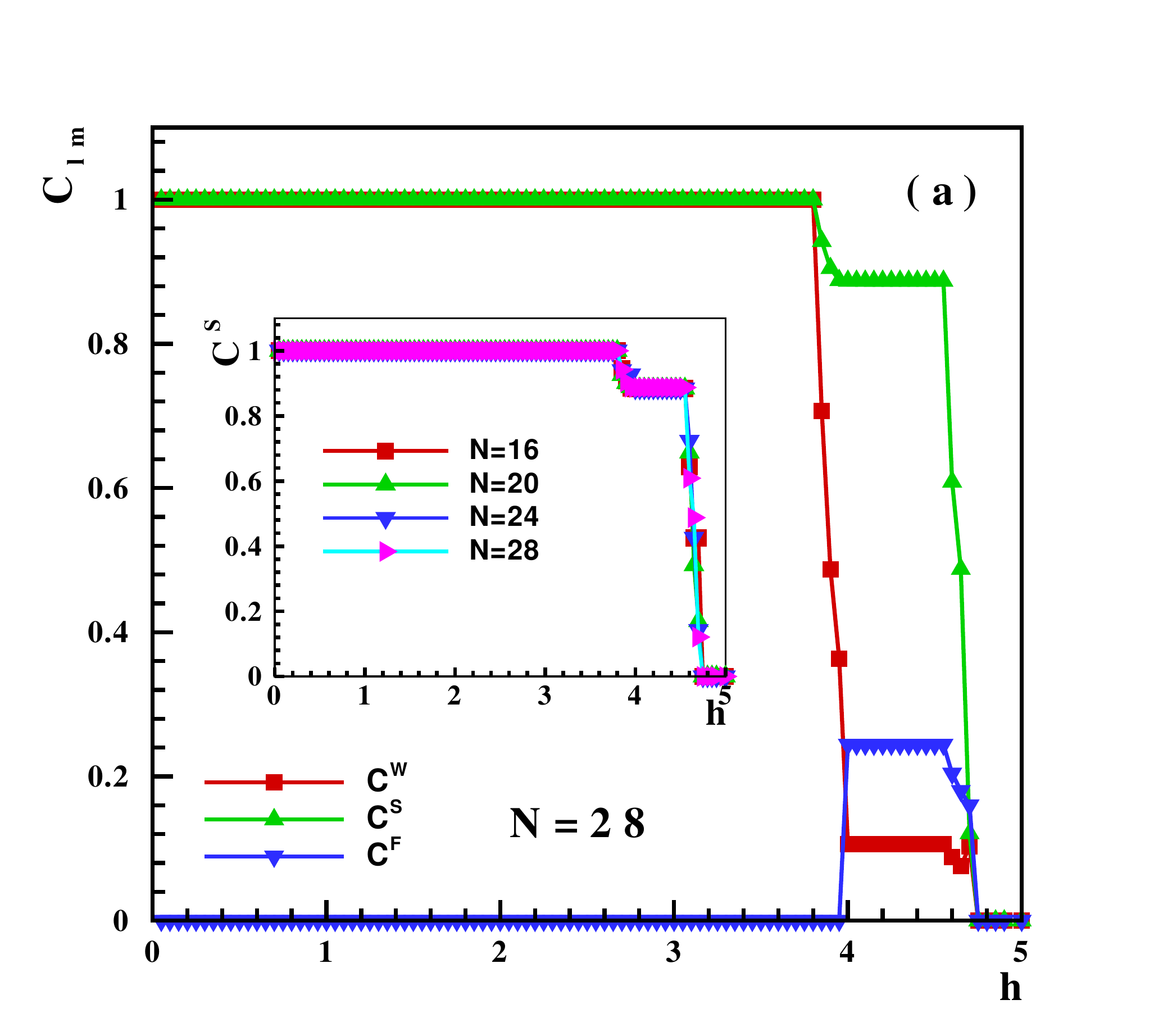}}
\centerline{\includegraphics[width=8cm,angle=0]{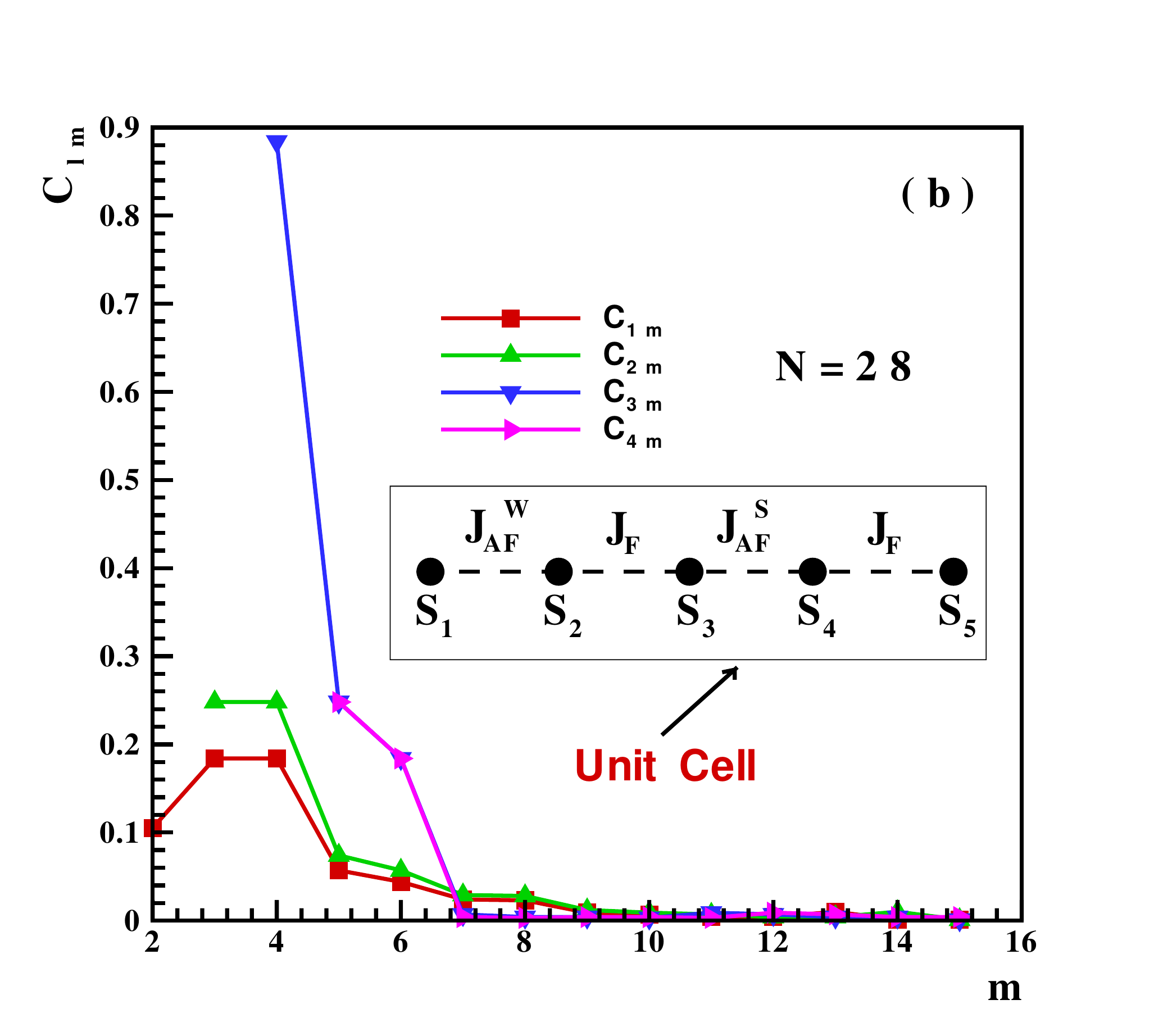}}
\caption{(Color online.) (a)-The concurrence as a function of the
applied field $h$, for chains with  exchanges $J_{F}=1.0$ ,
$J_{AF}=9/2$, and $\delta=1/9$ and length $N=28$. Inset plot: The
concurrence between two spins on strong bond as a
function of $h$ for different chain sizes $N=16$, $20$, $24$, and $28$.
(b)- Concurrence of two spins versus separation distance $m$.}
\label{concurrence-1}
\end{figure}
The numerical Lanczos results on the concurrence for the 1D
tetrameric spin-1/2 model have been shown in Fig.~(\ref{concurrence-1}).
We have plotted the entanglement of two spins
which are located at the same strong, weak, and ferromagnetic bond versus $h$, chain
length $N=28$ and the exchange parameters $J_{F}=1.0$, $J_{AF}=9/2$, and $\delta=1/9$.
In the Haldane phase, $h<h_{c_1}$, the spins on all antiferromagnetic
bonds make a singlet state. In this state which is a maximally entangled state,
$g_{lm}^{xx}=g_{lm}^{yy}=g_{lm}^{zz}=\frac{-1}{4}$, $C^S=C^W=1$, and $C^F$ is zero. For $h>h_{c_1}$ the values of $C^S$ and $C^W$ fall down with increasing of the magnetic field. Indeed the quantum correlations of the two spins with strong antiferromagnetic (SSA) and weak antiferromagnetic (SWA) interactions are decreased by increasing the magnetic field. However an enhancement on the entanglement of the two spins with ferromagnetic interaction (SF) is observed. It means that the magnetic field increases the quantum correlations of the two spins which are interacting ferromagnetically. This is a dual effect of the magnetic field in which increases the quantum correlations of two SF and decreases the quantum correlations of two SWA and two SSA.
In the intermediate gapless region $h_{c_1}<h<h_{c}^{-}$, the quantum correlations of SWA and SSA diminish down to $h=h_{c}^{-}$ and the concurrences $C^W$ and $C^S$ reduce to $\sim 0.1$ and $\sim 0.9$ respectively. However, the quantum correlations of SF grows up to the critical field $h_{c_-}$ and the entanglement reach to the value $\sim 0.25$. At the plateau state, the gap of the system is re-opened and a plateau emerges in the curve of concurrences. In the intermediate gapped phase the values of the concurrences $C^S$, $C^W$, and $C^F$ are $0.9$, $0.1$, and $0.25$, respectively.
Indeed in the plateau state there are three types of quantum correlations in the system.
These correlators are the source of the mid-plateaus in the different parameters of the system such as magnetization, bond dimer and F-dimer parameters.
Indeed all of these quantum correlators are exist only at the mid-plateau states. In the full saturated state all of them disappear and the entanglement of the state is exactly zero.

To see the finite size effects, we have plotted in the
inset plot of Fig.~\ref{concurrence-1}(a), the concurrence between two
SSA as a function of $h$ for different
chain sizes $N=16$, $20$, $24$, and $28$. It can be seen that there is not
size effect on the numerical results and the concurrence behaves
as thermodynamic limit in the gapped regions.

To get more intuition on the mid-plateau state, we have computed
numerically the entanglement between the spins with different
separation distances. The concurrence between a SSA and an
arbitrary spin, say $S_m$ has been  plotted versus the separation
distance $m$, in Fig.~\ref{concurrence-1}(b). The entanglement of
two such spins is decreased by increasing $m$ and goes to zero at
the finite value of $m=7$. The same behavior is also observed for
the entanglement between a SWA and the spin $S_m$ with respect to
$m$. The entanglement between SWA and $S_m$ goes to zero at $m=9$.
It means that, in the plateau state the range of the quantum
correlations between a SWA and $S_m$ is longer than the range of a
SSA and $S_m$. This is the other feature of the mid-plateau state.

It is also remarkable that from our numerical results we found that the
concurrence of two spins that are not on the same bond is equal to
zero in the gapped Haldane and saturated ferromagnetic phases
which is in complete agreement with the analytical results.


\section{conclusion}

In this paper, we have focused on the magnetic properties
of a bond-alternating antiferromagnetic-ferromagnetic spin-1/2 Heisenberg chain.
Using two analytical approaches and a numerical method we have studied
the effects of an external magnetic field and a space modulation on the
ground state properties of the system.
In the limit where the AF couplings are dominant, we mapped the model
(\ref{hamiltoni}) to an effective XXZ Heisenberg chain with
anisotropy parameter $\Delta=1/2$ in the presence of effective
uniform and staggered longitudinal magnetic fields. This model has
different quantum phases which are distinguished by four critical
fields. To find the critical fields we employed two field theoretical approaches
such as nonlinear sigma model and bosonization, and the numerical exact daigonalization Lanczos method.
Working on the spin coherent states representation, we mapped the model (\ref{hamiltoni}) to a nonlinear sigma model with an additional
topological term. The topological term is dependent on the space modulation $\delta$ and the parameter $\alpha=\frac{J_F}{J_{AF}}$.
For any values of $\alpha$ and $\delta$, the Hamiltonian mapped to a non-integrable
$O(3)$ NL$\sigma$M with $\Theta$ in the interval $[0,\pi]$. In the absence of a magnetic field the model is always
gapped and the gap value depends to $\alpha$ and
$\delta$. By analyzing our NL$\sigma$M in presence of
the magnetic field, we obtained only the two critical fields. To dominant this vacancy and to
find the other two critical fields we
also bosonized the effective Hamiltonian of
XXZ chain (\ref{effective-hamiltoni}). Our bosonization procedure showed that
the width of the mid-plateau is a function of $\delta J_{AF}$. It has been shown that the plateau-width, scales as a
power low with exponent value $2$ and vanishes at $\delta=0$.

Moreover, we implemented the Lanczos method to numerically
diagonalize chains with finite length up to $N=28$. Using the
exact diagonalization technique,  we calculated the energy gap,
magnetization, the string, the F-dimer, and the bond dimer order parameters and various
correlation functions for different values of the external
magnetic field. In good qualitative agreement with our analytical
results, we showed clearly that a space modulation on the
antiferromagnetic exchanges leads to generation of a gap in the
excitation spectrum of the system and correspondingly a magnetization mid-plateau at $\frac{M}{M_{sat.}}=\frac12$.
We found that a non-zero plateau also creates in the plot of F-dimer, bond dimer order parameters.

To get more physical insight on the mid-plateau state we also investigated the pair-wise entanglement between
two different spins of the system. As a measure of entanglement the concurrence
between two arbitrary spins computed as a function of magnetic field.
A plateau is also appeared in the concurrence at the middle gapped state.
In the plateau state there are three types of quantum correlations in the system.
in the plateau state the range of the quantum
correlations between a spin on the weak antiferromagnetic bond and a $S_m$ is longer than the range of the quantum correlations of a
spin on the strong antiferromagnetic bond and $S_m$.
\section{Acknowledgments}

JA thanks A. Langari for his fruitful suggestions on the
manuscript. We are grateful to G. I. Japaridze and T. Vekua for their useful
comments.


\end{document}